\documentclass[twoside,fleqn]{ActaStyle}
\usepackage{times}
\usepackage{latexsym}
\usepackage{lscape}
\usepackage[tbtags]{amsmath}

\def\authorheading{D. Anselmi}

\def\shorttitle{Irreversibility of the RG flow}
\begin{document}
\pagerange{1}{8}

\title{SUM RULES FOR TRACE ANOMALIES AND IRREVERSIBILITY OF THE
RENORMALIZATION-GROUP FLOW}

\author{Damiano Anselmi}{Dipartimento di Fisica Enrico Fermi, Universit\`a 
di Pisa, and INFN}

\abstract{I review my explanation of the irreversibility of the
renormalization-group flow in even dimensions greater than two and address
new investigations and tests.}


The purpose of my talk is to review exact results for the trace anomalies $c$ and $a$ in
the critical limits of supersymmetric theories, which provide
non-perturbative evidence that the renormalization-group (RG)\
flow is irreversible, and then derive universal sum rules 
for $c$, $a$ and $a^{\prime }$ in even dimensions, formulate a theory of the
irreversibility of the RG flow, recapitulate the arguments and the evidence in favor
of this theory, and address new investigations and tests
of the predictions of this theory, analytical and on the lattice.

I consider the most general renormalizable quantum field theory, such that
the RG flow interpolates between UV and IR conformal fixed points. The
theory is embedded in external gravity. Let $\Gamma [g_{\mu \nu }]$ denote
the quantum action in the gravitational background. The trace anomaly is
given by the derivative of $\Gamma $ with respect to the conformal factor of
the metric:
\begin{equation}
\Theta =-\left. \frac{\delta \Gamma [g_{\mu \nu }{\rm e}^{2\phi }]}{\delta
\phi }\right| _{\phi =0}.  \label{deri}
\end{equation}
At criticality, the trace anomaly has the form
\begin{equation}
\Theta =\frac{1}{120}{\frac{1}{(4\pi )^{2}}}\left[ c\,W^{2}-{\frac{a}{4}}%
\,{\rm G}+{\frac{2}{3}}\,a^{\prime }\,\Box R\right] ,  \label{teta*}
\end{equation}
and the constants $c$, $a$ and $a^{\prime }$ are called central charges.
Here $W$ is the Weyl tensor ($W^{2}=R_{\mu \nu \rho \sigma }R^{\mu \nu \rho
\sigma }-2R_{\mu \nu }R^{\mu \nu }+\frac{1}{3}R^{2}$) and ${\rm G}=4R_{\mu
\nu \rho \sigma }R^{\mu \nu \rho \sigma }-16R_{\mu \nu }R^{\mu \nu }+4R^{2}$
is the Euler density.
In a free-field theory of $n_{s}$ real scalars, $n_{f}$ Dirac fermions and $%
n_{v}$ vectors, the values of $c$ and $a$ are
\begin{equation}
c=n_{s}+6n_{f}+12n_{v},\qquad a=\frac{1}{3}\left(
n_{s}+11n_{f}+62n_{v}\right) .  \label{uvv}
\end{equation}
The quantity $a^{\prime }$, instead, has an ambiguity. The quantum action is
defined up to the addition of arbitrary, finite local terms. The scalars of
dimension four constructed with the curvature tensors and their covariant
derivatives are $W^{2},$ ${\rm G,}$ $\Box R$ and $R^{2}$. $\int \sqrt{%
g}W^{2}$ gives zero in (\ref{deri}). $\int \sqrt{g}\Box R$ trivially
vanishes. $\sqrt{g}{\rm G}$ is a total derivative in four dimensions and 
so $%
\int \sqrt{g}{\rm G}$ does not contribute. There remains $\int \sqrt{g}R^{2}$%
. Using (\ref{deri}) we see that
\begin{equation}
\!\!\!\!\!\!\!\!\!\!{\Gamma \rightarrow \Gamma +{\frac{1}{2160}}\frac{\delta a^{\prime }}{(4\pi
)^{2}}\int \sqrt{g}R^{2}\quad \Rightarrow \quad \Theta \rightarrow \Theta +{%
\frac{1}{180}}\frac{\delta a^{\prime }}{(4\pi )^{2}}\Box R\quad \Rightarrow
\quad a^{\prime }\rightarrow a^{\prime }+\delta a^{\prime }.}  \label{amb}
\end{equation}
The ambiguity $\delta a^{\prime }$ is RG invariant, since the local terms
must be finite. Therefore, $\delta a^{\prime }$ cancels out in differences
such as $\Delta a^{\prime }=a_{{\rm UV}}^{\prime }-a_{{\rm IR}}^{\prime }$.
Nevertheless, while $\Delta c$ and $\Delta a$ are flow invariants, i.e. they do not depend on the
particular flow connecting two fixed points, $\Delta a'$ is not a flow invariant \cite{inv}.

At intermediate energies, the coefficients $c$, $a$ and $a^{\prime }$ depend
on the coupling constant $\alpha $ (and the subtraction scheme). For
example, in (massless) QED\ or QCD, the trace anomaly operator equation has
the form
\[
\Theta =\frac{1}{120}\frac{1}{(4\pi )^{2}}\left[ c(\alpha )\ W^{2}-\frac{1}{%
4}a(\alpha )\ {\rm G}+\frac{2}{3}a^{\prime }(\alpha )\ \Box R+\beta (\alpha
)h(\alpha )\ R^{2}\right] -\frac{1}{4}\ \beta (\alpha )F^{2},
\]
where $\beta (\alpha )={\rm d\ln }\alpha /{\rm d\ln }\mu $, $h(\alpha )$ in
an unspecified, regular function of $\alpha $ and $F$ is the field strength
of the gauge field. In flat space, $\Theta =-\beta (\alpha )F^{2}/4$.

The central charges are universally normalized as follows: $c$ is
normalized to be 1 for the free real scalar field; the relative
normalization of $c$ and $a$ is fixed in such a way that the
subclass of renormalization-group flows with $\Delta c=\Delta a$
is special in a sense to be specified below; the relative
normalization of $a^{\prime }$ and $a$ is fixed in such a way that
the subclass of renormalization-group flows with $\Delta a^{\prime
}=\Delta a$ is special in another sense to be explained below.

Exact results for $c$ and $a$ in interacting conformal field theories are
known \cite{noi,noi2}. For example, N=1 ${\rm SU}(N_{c})$ supersymmetric QCD
with $N_{f}$ massless quark flavors in the conformal window
$3N_{c}/2\leq N_{f}\leq 3N_{c}$, has an interacting fixed point in the infrared, and
\[
c_{{\rm IR}}=\frac{15}{2}\left( 7N_{c}^{2}-2-9\frac{N_{c}^{4}}{N_{f}^{2}}%
\right) ,~~~~a_{{\rm IR}}=\frac{45}{2}\left( 2N_{c}^{2}-1-3\frac{N_{c}^{4}}{%
N_{f}^{2}}\right) .
\]
The IR values of the central charges $c$ and $a$ are derived combining the Adler-Bardeen
theorem with the relation between the trace and axial anomalies, provided by supersymmetry. Due
to the Adler-Bardeen theorem, a certain class of axial anomalies is one-loop
exact \cite{thooft}. The relation between the trace and axial anomalies
is such that, in most supersymmetric theories, exact results can
be obtained for the trace anomaly also, and one-loop calculations are enough
to extract $c_{{\rm IR}}$ and $a_{{\rm IR}}$.

Using the UV values (\ref{uvv}), we obtain the differences
\begin{eqnarray*}
\Delta c &=&c_{{\rm UV}}-c_{{\rm IR}}=\frac{5}{2}N_{c}N_{f}\left( 1-3\frac{%
N_{c}}{N_{f}}\right) \left( 4-3\frac{N_{c}}{N_{f}}-9\frac{N_{c}^{2}}{%
N_{f}^{2}}\right) , \\
\Delta a &=&a_{{\rm UV}}-a_{{\rm IR}}=\frac{5}{2}N_{c}N_{f}\left( 1-3\frac{%
N_{c}}{N_{f}}\right) ^{2}\left( 2+3\frac{N_{c}}{N_{f}}\right) .
\end{eqnarray*}
It is immediate to verify that, in the conformal window, $a_{%
{\rm UV}}\geq a_{{\rm IR}}\geq 0$ and $c_{{\rm IR}}\geq 0$, while $\Delta c$
can be either positive or negative. The inequalities $a_{{\rm UV}}\geq a_{%
{\rm IR}}\geq 0$ and $c_{{\rm IR}}\geq 0$ are satisfied in all the models
studied in \cite{noi,noi2}.

The property $a_{{\rm UV}}\geq a_{{\rm IR}}$ $\geq 0$ means that there 
exists
a positive function $a$ which always decreases along the RG\ flow, from the
UV to the IR. This property is called irreversibility of the RG flow. In the
rest of the talk I present my theory of the irreversibility of the RG flow.

\medskip

Universal sum rules for the trace anomalies $c$, $a$ and $a^{\prime }$ can
be derived \cite{234}. For simplicity, we take a conformally flat metric
$g_{\mu \nu }=\delta _{\mu \nu }{\rm e}^{2\phi }$, 
which is enough to study
the correlation functions containing insertions of $\Theta $, since $\phi 
$ is the
external source coupled to the operator $\Theta $. The $\Theta $-correlators
have complicated expressions, at intermediate energies, but tend to
universal, simple limits at criticality, which contain
just two local structures, multiplied by $a$ and $a^{\prime }$, and can be
derived taking $\phi $-derivatives of (\ref{teta*}%
). The sum rules for $\Delta a=a_{{\rm UV}}-a_{{\rm IR}}$ and $\Delta
a^{\prime }=a_{{\rm UV}}^{\prime }-a_{{\rm IR}}^{\prime }$ are direct
consequences of this. Studying the critical limits of the $\Theta$-correlators
in the sense of distributions, we arrive at \cite{234}
\begin{eqnarray}
\Delta a^{\prime } &=&{\frac{5\pi ^{2}}{2}}\int {\rm d}^{4}x\,|x|^{4}\,%
\langle \widetilde{\Theta }(x)\,\widetilde{\Theta }(0)\rangle ,  \nonumber \\
\Delta a-\Delta a^{\prime } &=&{\frac{5\pi ^{2}}{2}}\int {\rm d}^{4}x\,{\rm d%
}^{4}y\,\,x^{2}\,y^{2}\,\left\{ \langle \widetilde{\Theta }(x)\,\widetilde{%
\Theta }(y)\,\widetilde{\Theta }(0)\rangle \right.  \label{tilde} \\
&&\left. \qquad \qquad \qquad \qquad \qquad +\left\langle {\frac{\widetilde{%
\delta }\widetilde{\Theta }(x)}{\widetilde{\delta }\phi (y)}}\,\widetilde{%
\Theta }(0)\right\rangle +2\left\langle {\frac{\widetilde{\delta }\widetilde{%
\Theta }(x)}{\widetilde{\delta }\phi (0)}}\,\widetilde{\Theta }%
(y)\right\rangle \right\} .  \nonumber
\end{eqnarray}
The difference $\Delta a-\Delta a^{\prime }$ can be expressed in many
equivalent ways \cite{234}. Here I\ have reported the simplest formula. Sum
rules for $\Delta c$ can be derived also, but they are more complicated.

In formulas (\ref{tilde}), the tildes mean that certain terms
proportional to the field equations have been eliminated. Denoting
with $\varphi $ the
dynamical fields of the theory, with conformal weight $h$, the notation $%
\widetilde{\delta }/\widetilde{\delta }\phi $ stands for the derivative with
respect to $\phi $ at constant $\widetilde{\varphi }\equiv \varphi \,{\rm e}%
^{h\phi }$. In particular, $\widetilde{\Theta }=-{\widetilde{\delta }S/%
\widetilde{\delta }\phi }$, where $S$ denotes the action. It is understood
that in (\ref{tilde}) $\phi $ is set to zero, after taking the $\phi $-derivatives of $\widetilde{\Theta }$. 

Osterwalder-Schrader (OS)\ positivity \cite{os} can be applied to the sum
rule for $\Delta a^{\prime }$, and implies $\Delta a^{\prime }\geq 0$. This
is not precisely what we want, since $a^{\prime }$ is ill-defined at criticality by the
addition of an arbitrary constant and $\Delta a'$ is not a flow invariant. OS positivity is ineffective on the sum
rules for $\Delta a$, so that it is not immediate to derive $\Delta a\geq 0$
from (\ref{tilde}).

\medskip

In \cite{athm,at6d}, I have proposed a solution to this puzzle, which I now
recall. An immediate observation is that the problem is solved in the class
of flows which satisfy
\begin{equation}
\Delta a\geq f~\Delta a^{\prime },  \label{expl1}
\end{equation}
where $f$ is a non-negative numerical factor. We would like to understand if
this class of theories is sufficiently interesting. The method of \cite
{noi,noi2} to compute $\Delta a$ and $\Delta c$ in supersymmetric theories
does not allow us to compute $\Delta a^{\prime }$, so we have to search for
new arguments. It is helpful to treat, at a time, the case
of generic even dimension $d=2n$, so as to include, in particular,
two-dimensional quantum field theory, which satisfies Zamolodchikov's $c$%
-theorem \cite{zamolo}. The trace anomaly at criticality in even dimensions
contains three types of terms constructed with the curvature tensors and their
covariant derivatives: {\it i}) terms ${\cal W}_{i}$, $i=0,1,\ldots
,I$, such that $\sqrt{g}{\cal W}_{i}$ are conformally invariant; {\it ii})
the Euler density
\[
{\rm G}_{d}=(-1)^{n}\varepsilon _{\mu _{1}\nu _{1}\cdots \mu _{n}\nu
_{n}}\varepsilon ^{\alpha _{1}\beta _{1}\cdots \alpha _{n}\beta
_{n}}\prod_{i=1}^{n}R_{\alpha _{i}\beta _{i}}^{\mu _{i}\nu _{i}}~;
\]
{\it iii}) covariant total derivatives ${\cal D}_{j}$, $j=0,1,\ldots ,J$,
having the form $\nabla _{\alpha }J^{\alpha }$, $J^{\alpha }$ denoting a
covariant current. The coefficients multiplying these terms in the
trace anomaly are denoted with $c_{d}^{i}$, $a_{d}$ and $a_{d}^{j~\prime }$, respectively.
We write $c_{d}=c_{d}^{0}$ and $a_{d}^{\prime }=a_{d}^{0~\prime }$. We have\[
\Theta _{d{\rm =2}}^{*}=\frac{1}{24\pi }c_{2}R,\qquad \qquad \Theta _{d{\rm %
=4}}^{*}=\frac{1}{120}{\frac{1}{(4\pi )^{2}}}\left[ c_{4}\,W^{2}-{\frac{a_{4}%
}{4}}\,{\rm G}_{4}+{\frac{2}{3}}\,a_{4}^{\prime }\,\Box R\right] ,
\]
\begin{equation}
\!\!\!\!\!\!\!\!\!\!\!\!{\Theta _{d{\rm =}2n}^{*}={\frac{{n}!}{(4\pi )^{n}\,(d+1)!}}\left[ {\frac{%
c_{d}\left( d-2\right) }{4(d-3)}}{\cal W}_{0}+\sum_{i=1}^{I}c_{d}^{i}{\cal W}%
_{i}-{\frac{2^{1-n}}{\,d}}\left( a_{d}{\rm G}_{d}+\sum_{j=0}^{J}a_{d}^{j~%
\prime }{\cal D}_{j}\right) \right] .}  \label{trace}
\end{equation}
Here $c_{2}$ is $a_{2}$.
${\cal W}_{0}$ is the unique term of the form $W\Box ^{n-2}W+\cdots $
such that $\sqrt{g}{\cal W}_{0}$ is conformally invariant, where the dots
denote cubic terms in the curvature tensors. We choose a basis in which 
the ${\cal W}_{i}$s with $i=1,\ldots,I$ are at least cubic in the 
curvature tensors. I have  separated ${\cal W}_{0}$
from the other ${\cal W}_{i}$s, because its coefficient $%
c_{d}$ is also the coefficient of the stress-tensor two-point function (see \cite{cea} for details)
%\[
%\langle T_{\mu \nu }(x)\,T_{\rho \sigma }(0)\rangle =c_{d}{\frac{n!(n-1)!}{%
%(2\pi )^{d}(d+1)!}}{\prod }_{\mu \nu ,\rho \sigma }^{(2)}\Box ^{n-2}\left( {%
%\frac{1}{|x|^{d}}}\right) ,
%\]
and is normalized so that for free fields ($n_{s}$ real scalars, $n_{f}$
Dirac fermions and $n_{v}$ ($n-1$)-forms) it reads
\begin{equation}
c_{d}=n_{s}+2^{n-1}(d-1)n_{f}+\frac{d!}{2\left[ \left( n-1\right) !\right]
^{2}}n_{v}.  \label{freec}
\end{equation}
The covariant total derivative term
\[
{\cal D}_{0}=-{\frac{2^{n}d}{2(d-1)}}\Box ^{n-1}R,
\]
needs to be singled out among the ${\cal D}_{j}$s, since it is
the unique term linear in the curvature tensors. We choose a basis such that
all the ${\cal D}_{j}$s, $j>0$, are at least quadratic in the curvature
tensors. Then, on conformally flat metrics, ${\cal D}_{0}$ contains the
unique term linear in $\phi $. Proceeding as for the first formula of (\ref
{tilde}), the sum rule
\begin{equation}
\Delta a_{d}^{\prime }=\frac{\pi ^{n}\,(d+1)}{{n}!}\int {\rm d}%
^{d}x\,|x|^{d}\,\langle \Theta (x)\,\Theta (0)\rangle   \label{sum2}
\end{equation}
is derived \cite{cea}, so that $\Delta a_{d}^{\prime }\geq 0$ in
arbitrary even dimensions.

The covariant total derivatives ${\cal D}_{j}$ are in one-to-one
correspondence with the arbitrary finite local terms that
can be added to $\Gamma $, and the coefficients $a_{d}^{j~\prime
}$ have
ambiguities similar to the ambiguity (\ref{amb}) of the coefficient $%
a^{\prime }$ in four dimensions. In \cite{at6d} the ${\cal W}_i$s and the ${\cal D}_j$s are studied
explicitly in $d=6$ and the ${\cal D}_j$s in $d=8$.

We first observe that, in two dimensions, on conformally flat metrics, the
trace anomaly at criticality is linear in the conformal factor,
\begin{equation}
\sqrt{g}\Theta _{d{\rm =2}}^{*}=\frac{1}{24\pi }c_{2}\sqrt{g}R=-\frac{1}{%
12\pi }c_{2}\Box \phi .  \label{d2}
\end{equation}
It can be shown that Zamolodchikov's $c$-theorem follows directly from this fact.
Indeed, in two dimensions $c_{2}$ plays also the role of $a_{2}^{\prime }$,
for which a sum rule similar to the first of (\ref{tilde}) can be derived
\cite{cardy}, implying $\Delta c_{2}\geq 0$. It is interesting to
investigate under which conditions $\sqrt{g}\Theta ^{*}$ is linear in $\phi$ on conformally flat metrics 
in arbitrary even dimension. We then discover \cite{at6d}
that there exists a ``pondered'' modification of the Euler density
\begin{equation}
\tilde{{\rm G}}_{d}={\rm G}_{d}-{\frac{2^{n}d}{2(d-1)}}f_{d}^{0}\Box
^{n-1}R+\sum_{j=1}^{J}f_{d}^{j}{\cal D}_{j}={\rm G}_{d}+\nabla _{\alpha
}J_{d}^{\alpha },  \label{pond}
\end{equation}
such that on conformally flat metrics
\[
\sqrt{g}\tilde{{\rm G}}_{d}=2^{n}d~f_{d}^{0}\Box ^{n}\phi .
\]
To fix the weight $f_{d}^{0}$, we compute the Euler characteristics of a $d$%
-dimensional sphere $S^{d}$ (equal to 2) with $\tilde{{\rm G}}_{d}$ and $%
{\rm G}_{d}$, and match the results \cite{at6d}. I have distributed
appropriate numerical factors, so far, such that the outcome of this calculation is $f_{d}^{0}=1.$ This
procedure fixes the relative normalization of $a$ and $a^{\prime }$. More
involved calculations are necessary to fix the other $f_{d}^{j}$s. In \cite
{at6d} complete expressions of $\tilde{{\rm G}}_{d}$ are written in $d=6$
and $d=8.$

On conformally flat metrics, the terms ${\cal W}_{i}$ drop and the trace
anomaly has the form
\begin{equation}
\Theta _{d=2n}^{*}=-N_{d}a_{d}^{*}\,\tilde{{\rm 
G}}_{d}-N_{d}\sum_{j=0}^{J}%
\left( a_{d}^{j*~\prime }-a_{d}^{*}f_{d}^{j}\right) {\cal D}_{j},
\label{tetastar}
\end{equation}
where the overall (positive) numerical factor $N_{d}$ is still unspecified.
The convention for the relative normalization of $c$ with respect to $a$ and
$a^{\prime }$ will be fixed later (in ref.s \cite{at6d,cea} it was $N_{d}=1$%
). In eq. (\ref{tetastar}) the stars refer to the values at criticality.

Using the RG-invariant ambiguities $\delta a^{\prime }$ of the $a^{\prime 
}s,$
and the existence of $\tilde{{\rm G}}_{d}$ (\ref{pond}), we can arrange the $%
a^{\prime }$s so that at one critical point, say the UV, the trace anomaly
is linear in $\phi $ on conformally flat metrics, namely it has the form
\begin{equation}
\sqrt{g}\Theta _{d=2n}^{{\rm UV}}=-N_{d}a_{d}^{{\rm 
UV}}\,\sqrt{g}\tilde{{\rm %
G}}_{d}=-2^{n}dN_{d}\,a_{d}^{{\rm UV}}\,\Box ^{n}\phi ,  \label{uv}
\end{equation}
precisely as in (\ref{d2}). This means, in particular, $a_{d}^{\prime \ {\rm %
UV}}=a_{d}^{{\rm UV}}$. Once this choice is made, however, the $a^{\prime 
}$s
are fixed at all energies. In particular, it is not obvious that the trace
anomaly has the form (\ref{uv}) also at the IR\ fixed point, i.e.
\begin{equation}
\sqrt{g}\Theta _{d=2n}^{{\rm IR}}=-N_{d}a_{d}^{{\rm 
IR}}\,\sqrt{g}\tilde{{\rm %
G}}_{d}=-2^{n}dN_{d}\,a_{d}^{{\rm IR}}\,\Box ^{n}\phi ,  \label{ir}
\end{equation}
nor that $a_{d}^{\prime \ {\rm IR}}=a_{d}^{{\rm IR}}$. Let us assume that this happens. Then,
we have the equality in (\ref{expl1}) with $f=1$.
The $\Delta a_{d}^{\prime }$-sum rule (\ref{sum2}) is promoted to a $\Delta
a_{d}$-sum rule and the RG flow is irreversible. Since the normalization factor $N_{d}$ is still unfixed, we find
\begin{equation}
\Delta a_{d}=\Delta a'_{d}=\frac{1}{2^{{3n}-1}\,d\,\Gamma (d+1)N_{d}}\int {\rm d}%
^{d}x\,|x|^{d}\,\langle \Theta (x)\,\Theta (0)\rangle .  \label{suma}
\end{equation}

\medskip

We have to understand when $a_{d}$ and $a_{d}^{\prime }$ can be identified
at both critical points of the RG\ flow, i.e. when $a_{d}^{\prime \ {\rm %
UV}}=a_{d}^{{\rm UV}}$ implies $a_{d}^{\prime \ {\rm %
IR}}=a_{d}^{{\rm IR}}$. I begin listing a few facts.

-- Explicit calculations show that free massive fields have $\Delta a_d\neq
\Delta a^{\prime }_d$. For example, the free massive scalar and the free
massive Dirac fermion in $d=4$ have
\[
\Delta a_{{\rm scal}}=\frac{1}{3},\qquad \Delta a_{{\rm scal}}^{\prime
}=\Delta c_{{\rm scal}}=1,\qquad \Delta a_{{\rm ferm}}=\frac{11}{3},\qquad
\Delta a_{{\rm ferm}}^{\prime }=\Delta c_{{\rm ferm}}=6.
\]
We see that $\Delta a^{\prime }$ and $\Delta c$ coincide in these cases. We
will have more to see about this coincidence later.

-- In perturbation theory, $\Delta a-\Delta a^{\prime }=0$ to
some non-trivial loop orders in classically conformal theories in $d=4$ \cite
{athm} and $d=6$ \cite{at6d}.

It is therefore necessary to distinguish two main classes of quantum field
theories:

{\it i}) the classically conformal theories. They contain only marginal
deformations, that is to say no dimensionful parameter other than the dynamically
generated scale, which I denote with $\mu $. Massless QCD belongs to this
class, as well as the conformal windows of \cite{noi}. The RG\ flow is
``pure'', i.e. not contaminated by the effects of classical scales.

{\it ii}) theories which are not conformal at the classical level,
because they contain relevant deformations, e.g. masses.

I claim that $\Delta a-\Delta
a^{\prime }$ is exactly zero in classically conformal theories, i.e.
that the pure RG flow is irreversible and satisfies (\ref{suma}). In \cite{athm,at6d} I showed
that this property can be derived from the following statement about the dependence
of the quantum action $\Gamma$ on the conformal factor:

{\bf S.} {\it The quantum action for the conformal factor }$\Gamma [\phi ]$ {\it is
bounded from below (in the Euclidean framework) at all energies if it is
bounded from below at some energy.}

The proof that {\bf S} implies $\Delta a=\Delta a^{\prime }$
in classically conformal theories can be found in \cite{athm} and is not
repeated here for reasons of space.
 
{\bf S} is suggested by the following considerations:

-- A statement like {\bf S} holds for the dependence of $\Gamma $ on the dynamical fields. This is
the requirement that the quantum action have a minimum in the space of
physical fields.

-- A statement like {\bf S} does not hold, in general, for the dependence of $\Gamma $ on external
sources, because divergences generate arbitrarily negative terms. These
terms can be normally reabsorbed in the running coupling constants, but when
the sources are external there are no such constants (or need to be
introduced anew).

-- The unique known case in which $\Gamma $ is convergent even in the
presence of external sources is when the external source is the conformal factor $\phi$ and the theory is 
classically conformal, because $\Theta $ is an evanescent operator. The rigorous derivation of
the convergence of $\Gamma [\phi ]\ $at arbitrary energies is in \cite{234}. We conclude that {\bf S} is
expected to hold for $\Gamma [\phi ]$ in classically conformal theories.

\medskip

Let us summarize the evidence in favor of this theory of the irreversibility of the RG flow. 
It explains the distinction between the two classes $i$) and $ii$) mentioned
above: $\Gamma [\phi ]$ is not convergent in the presence of masses. It
predicts the existence of the ``pondered'' Euler density $\tilde{{\rm 
G}}_{d}$ (\ref{pond}) and explains
its physical meaning. The predictions of this theory agree with the
perturbative calculations made so far, to four loops in $d=4$ and $d=6$. 
There exists a physical argument for the validity of this theory to all 
orders. It predicts and explains the existence of a remarkable class of theories, 
having $c=a$ (see
below), singled out also by independent arguments. I conclude with the
discussion of this last point. 

In two dimensions, there is no distinction between the classes $i$) and $ii$%
).\ This suggests that in even dimensions greater than two there exists a
subclass of flows where there is no distinction $i$)--$ii$) either.
This subclass, in particular, should have $\Delta a_{d}=\Delta a_{d}^{\prime
}$ and contain also theories with masses and classical scales.

To figure out how this subclass looks like, let us study the simplest
properties of massive flows. In particular, if we compute the integral of (%
\ref{suma}) for free massive scalars and fermions in arbitrary dimension, we
discover that the ratio between the values of the integral for scalars and
fermions equals the ratio between the scalar- and fermion-values of 
$c_{d}$
in (\ref{freec}):
\[
\int {\rm d}^{d}x\,|x|^{d}\,\langle \Theta (x)\,\Theta (0)\rangle ={\frac{%
\Delta c_{d}\,{n}!}{\pi ^{n}\,(d+1)}},\qquad {\rm i.e.\quad }\Delta c_{d}=%
\frac{2^{{3n}-1}}{{n}!}\pi ^{n}\,d\,(d+1)!N_{d}\Delta a_{d}^{\prime }{\rm .}
\]
This fact is known to hold by explicit computation, but 
a complete understanding is still lacking: see \cite{inv} for more details. We conclude
that there exists a class of classically nonconformal theories where $%
\Delta c_{d}$ is related to $\Delta a_{d}^{\prime }$. We can fix the
normalization factor $N_{d}$ such that this class has precisely $\Delta
c_{d}=\Delta a_{d}^{\prime }$, i.e.
\[
N_{d}=\frac{{n}!}{2^{{3n}-1}\pi ^{n}\,d\,(d+1)!}.
\]

\noindent Recapitulating, we have identified two main classes of
flows, so far:

1) the flows which have $\Delta a_{d}=\Delta a_{d}^{\prime }$;

2) the flows which have $\Delta c_{d}=\Delta a_{d}^{\prime }$.

\noindent We have seen that the theories $i$) satisfy 1) and that the flows
2) contain theories of $ii$). Therefore, we can argue that the subclass of
flows where classical conformality is violated but the equality $\Delta a_{d}=\Delta
a_{d}^{\prime }$ still holds, is the class of flows with $\Delta c_{d}=\Delta a_{d}$. In
particular, we have $\Delta c_{d}=\Delta a_{d}$ when both the UV\ and IR\
fixed points have $c_{d}=a_{d}$. If we plug this relation in the expression (%
\ref{trace}) of the trace anomaly at criticality, we single out the combination
\begin{equation}
{\cal G}_{d}={\rm G}_{d}-{\frac{2^{{n}-3}d(d-2)}{d-3}}{\cal W}%
_{0}+\sum_{i=1}^{I}h_{d}^{j}{\cal W}_{i}{\rm ,}  \label{comba}
\end{equation}
where $h_{d}^{j}$ are yet-unspecified numerical factors. The conformal field theories with
$c_d=a_d$ have a trace
anomaly proportional to ${\cal G}_{d}$ up to the ${\cal
D}_{j}$s.
The combination ${\cal G}_d$ is
also known from arguments completely independent of our
considerations about the irreversibility of the RG flow. First, in $d=4$ we have
\[
{\cal G}_{4}={\rm G}_{4}-4W^{2}=-8R_{\mu \nu }^{2}+\frac{8}{3}R^{2}.
\]
The square of the Riemann tensor drops out in the difference.
An example of theory with $c=a$ in $d=4$ is $N=4$ supersymmetric Yang-Mills theory, whose
peculiarity is well-known. The combination
(\ref{comba}) in six dimensions was pointed out in \cite{bonora} and in arbitrary
even dimensions by the authors of \cite {skende}. Actually, 
the construction of \cite
{bonora,skende} fixes uniquely also the $h_{d}^{i}$s. Finally, the structure of $%
{\cal G}_d$ is uniquely characterized as pointed out in 
\cite{cea}, namely ${\cal G}_d$ has the form
\[
{\cal G}_{d}=R_{\mu \nu }{\cal T}_{d}^{\mu \nu \rho \sigma }R_{\rho \sigma }%
{\rm ,}
\]
where ${\cal T}_{d}^{\mu \nu \rho \sigma }$ is a local covariant
differential operator of dimension $d-4$, constructed with the curvature
tensors, their covariant derivatives, and the covariant-derivative operator $%
\nabla $, acting both on the left and on the right.

\medskip

To conclude, I have formulated a theory of the irreversibility of the RG flow which
leads to the claim that
\begin{equation}
\Delta a_{d}=\frac{\pi ^{n}\,(d+1)}{{n}!}\int {\rm d}^{d}x\,|x|^{d}\,\langle
\Theta (x)\,\Theta (0)\rangle \label{magna}
\end{equation}
in the classically conformal quantum field theories and in the flows with $\Delta
c_{d}=\Delta a_{d}$. Irreversibility ($\Delta a_{d}\geq 0$) is then a consequence
of reflection positivity. The sum rules (\ref{tilde}) and (\ref{magna}) apply 
also to non-unitary flows \cite{inv,impr}. An appealing 
graphical interpretation of 
formula (\ref{magna}) is that $\Delta a_{d}$ is proportional to the 
scheme-invariant area of the graph of the beta function between the 
fixed points \cite{athm}.

As a byproduct, I have found a number of other results
which are important steps towards the classification of all
conformal theories and renormalization-group flows in arbitrary even
dimensions greater than two.

Further checks of the predictions of my theory can be made using the
universal sum rules (\ref{tilde}). One possibility is to test the
equality of $\Delta a$ and $\Delta a^{\prime }$ to the fifth or higher loop
orders in four-dimensional classically conformal theories, using the second
sum rule of (\ref{tilde}). Work is in progress in this direction. Another
possibility is to study the formulas (\ref{tilde}) and (\ref{magna}) on the
lattice. This can provide knowledge about the low-energy limit of QCD. For
example, the existence of a mass gap in QCD could be tested computing the
difference $\Delta a=a_{{\rm UV}}-a_{{\rm IR}}$ using (\ref{tilde}) and
knowing that $a_{{\rm UV}}$ is given by (\ref{uvv}), while $a_{{\rm IR}}=0$.
In massless QCD, it should be
possible to test the expected, non-trivial value of $a_{{\rm IR}}$,
proportional to the number of pions, and so have an indirect check that
quarks and gluons generate (the right number of) pions at low energies.

\end{document}